\documentclass[conference]{IEEEtran}
\usepackage{cite}
\usepackage{amsmath,amssymb,amsfonts}
\usepackage{algorithmic}
\usepackage{graphicx}
\usepackage{textcomp}
\usepackage{xcolor}
\usepackage{comment}
\usepackage{enumitem}
\usepackage{tikz}
\usetikzlibrary{shapes, arrows.meta, positioning, calc}
\usepackage[margin=1in]{geometry}
\usepackage{booktabs}
\usepackage{textcomp}

\usepackage{hyperref}

\def\BibTeX{{\rm B\kern-.05em{\sc i\kern-.025em b}\kern-.08em
    T\kern-.1667em\lower.7ex\hbox{E}\kern-.125emX}}

\begin{document}

\title{Toward Personalizing Quantum Computing Education: An Evolutionary LLM-Powered Approach}

\author{\IEEEauthorblockN{Iizalaarab Elhaimeur}
\IEEEauthorblockA{\textit{Center for Real-Time Computing} \\
\textit{Computer Science Department}\\
\textit{Old Dominion University}\\
Norfolk, VA \\
ielha003@odu.edu}
\and
\IEEEauthorblockN{
Nikos Chrisochoides}
\IEEEauthorblockA{\textit{Center for Real-Time Computing} \\
\textit{Computer Science and Physics Departments}\\
\textit{Old Dominion University}\\
Norfolk, VA \\
nikos@cs.odu.edu}
}

\maketitle

\begin{abstract}

Quantum computing education faces significant challenges due to its complexity and the limitations of current tools; this paper introduces a novel Intelligent Teaching Assistant for quantum computing education and details its evolutionary design process.  The system combines a knowledge-graph-augmented architecture with two specialized Large Language Model (LLM) agents: a Teaching Agent for dynamic interaction, and a Lesson Planning Agent for lesson plan generation. The system is designed to adapt to individual student needs, with interactions meticulously tracked and stored in a knowledge graph.  This graph represents student actions, learning resources, and relationships, aiming to enable reasoning about effective learning pathways. We describe the implementation of the system, highlighting the challenges encountered and the solutions implemented, including introducing a dual-agent architecture where tasks are separated, all coordinated through a central knowledge graph that maintains system awareness, and a user-facing tag system intended to mitigate LLM hallucination and improve user control. Preliminary results illustrate the system's potential to capture rich interaction data, dynamically adapt lesson plans based on student feedback via a tag system in simulation, and facilitate context-aware tutoring through the integrated knowledge graph, though systematic evaluation is required.

\end{abstract}


\section{Introduction}

Quantum computing offers a revolutionary paradigm shift, but a significant workforce gap hinders its progress \cite{hughes2022assessing}. Teaching quantum computing is uniquely challenging, demanding an interdisciplinary understanding of physics, computer science, and mathematics, compounded by the counterintuitive nature of quantum principles. Traditional teaching methods and tools often fail, one of the many reasons is students’ diverse background \cite{PhysRevPhysEducRes.18.010150}. On the other hand, novel methods and tools based on generative artificial intelligence are still unproven in terms of successful teaching practices and quantifiable results. Many existing LLM-based tutors are like a conversational partner with amnesia – they can't recall what occurred just moments ago. They lack context-awareness, the ability to track a student’s progress, remember their questions, and use that information to personalize the learning experience. This, combined with the absence of robust planning, renders them insufficient for addressing the intricacies of education in general \cite{giannakos2024promise} and in our case, quantum computing training and upskilling ~\cite{Nita2020TheCA}. The need for personalized approaches is increasing \cite{chrisochoides2024developing}, particularly with the possible harms AI can introduce to learning if not carefully designed \cite{bastani2024generative}.

Current resources, including foundational textbooks like Nielsen and Chuang \cite{nielsen2010quantum} and online courses or interactive platforms, lack the necessary personalization and dynamic interaction of theory and practice \cite{mykhailova2020teaching}. While useful, resources like Microsoft’s Quantum Katas \cite{mykhailova2020quantum} typically have standardized content and follow predefined learning paths. Although LLMs show promise in education \cite{giannakos2024promise}, they are plagued by limitations in context-awareness and the ability to execute long-term planning \cite{lecun2022path}, \cite{wang2025llm}. Recognizing these critical shortcomings in personalization, dynamic theory-practice integration, and the current limitations of even promising LLM technologies, our work aims to address these three barriers directly.



Our approach aims to tackle these constraints through an iterative design, leading to a context-aware, graph-augmented, two-agent LLM system for personalized quantum computing instruction. We use two different LLM agents:

\begin{itemize}
    \item \textbf{Teaching Agent}: Manages real-time interaction, step-by-step instructions, and reacts to user-selected tags.
    \item \textbf{Lesson Planning Agent}: Develops and customizes lesson plans according to student requirements and teacher input.
\end{itemize}

A shared Knowledge Graph provides persistent memory, captures relationships, constructs learning pathways, and supports both agents. This approach offers context-awareness, personalized feedback, effective material access, and a basis for analysis of learning trajectories. End-user-selectable tags increase engagement and reduce LLM limitations.

\noindent
{\bf Contributions} This paper presents the evolutionary architecture of an intelligent teaching assistant system (ITAS) for quantum computing education. The key contributions are:

\begin{itemize}
    \item An innovative framework integrating two distinct LLM agents with a knowledge graph for context-aware tutoring structure.
    \item A comprehensive graph-based data model to capture student-system interactions and learning paths.
    \item Dynamic lesson execution with a separate lesson generation agent for flexible adaptation.
    \item A tag-based interaction system for improving user control and mitigating LLM hallucinations.
    \item An in-depth explanation of the iterative design process, highlighting problems and solutions.
\end{itemize}

The remainder of the paper is structured as follows.Section II reviews the related work. Section III presents the challenges of the quantum education platform and our solution. Section IV outlines the proposed system architecture,its development, along with certain initial findings. Section V outlines future work and closing remarks.

\section{Related Work}

While a detailed review of prior work in quantum education is beyond our current scope due to space limitations, we direct readers to ~\cite{wilcox_24}, \cite{juarez2024skills}, \cite{nita2023challenge}, \cite{dundar2025making}, and \cite{9705217} for relevant related articles.

Large Language Models (LLMs) are increasingly explored for educational purposes. Their abilities range from creating instructional materials, like lesson plans \cite{hu2024teaching}, to providing feedback and assessment \cite{prather2023robots}. But serious problems remain. Over-reliance on LLM dialogue systems can potentially impede the development of analytical and critical thinking skills \cite{zhaieffects}. Additionally, problems such as LLM hallucination, bias, and ensuring ethical use are foremost concerns \cite{giannakos2024promise}, \cite{prather2023robots}. Although there is vast potential, successful integration requires careful consideration of pedagogy and mitigation of risks.

Knowledge Graphs (KGs) are a structured method to represent domain knowledge and relationships, making them valuable for personalization in education \cite{qu2024survey}. Personal Knowledge Graphs (PKGs) specifically model individual learner states, interactions, and preferences, facilitating personalized learning experiences \cite{skjaeveland2024ecosystem}. Systems are being constructed that use PKGs to model learner knowledge based on their interactions and recommend appropriate instructional materials \cite{ain2024learner}. The combination of symbolic knowledge representation (KGs) with sub-symbolic neural models (LLMs), often termed Neurosymbolic AI, has the potential to generate robust and interpretable educational AI systems \cite{jaldi2025education}.

The Intelligent Tutoring Systems (ITS) aim to provide tailored instruction comparable to that of human tutors. Current ITS utilize AI, typically with conversational interfaces powered by LLMs \cite{chu2025llm}. Multi-agent systems, where different agents handle specific functions such as planning, teaching, or assessing, are a growing trend \cite{wang2025llm, zhao2025sirius, chu2025llm}. Such systems can be proactive, anticipating the needs of students instead of merely responding to queries \cite{lu2024proactive}. Particular applications such as Tutorly show how LLMs can be combined with video content to form apprenticeship-style learning environments \cite{li2024tutorly}. The goal is typically to develop adaptive systems that can enhance over time, potentially through self-improving mechanisms \cite{zhao2025sirius}, while maintaining pedagogical soundness \cite{jaldi2025education}.

Video is a popular medium in online education, from MOOCs to individual tutorials \cite{giannakos2014collecting, giannakos2015making}. Analyzing learner interaction data (clickstreams, pauses, replays) provides insights into engagement and confusion points \cite{giannakos2015making}. Learnersourcing, in which collective learner activity data is utilized for improving the video interface or content (e.g., inserting navigation aids, summaries, or identifying important segments), has emerged as a powerful technique \cite{kim2015learnersourcing}. This data-oriented approach can tailor the educational experience \cite{chrisochoides2024developing} and inform curricular development, such as in flipped classroom models, which often rely on pre-recorded video lectures \cite{giannakos2014reviewing}.

\section{Challenges in Existing LLM-Powered Quantum Education Platforms and Our Solutions}

This section examines limitations in state-of-the-art LLM-powered quantum education platforms, using Quantum Katas \cite{mykhailova2020quantum} as a representative example, and contrasts them with solutions incorporated into our system. 
It is estimated (by experts) that it may take 3-5 years for sufficient domain-specific data to be generated to allow for a single, general-purpose large language model to become truly effective in assisting and teaching quantum computing without major architectural support. This setting emphasizes the necessity for well-thought-out systems today. 

In this paper, we detail the design and improvement of our Intelligent Teaching Assistant System (ITAS) architecture. Based on lessons learned from current platforms, the design aims to overcome key observed limitations by specifically addressing three challenges in current approaches: (i) orthogonality, (ii) reliance on static learning pathways, and (iii) potential for student over-dependence on large language models (LLMs). Subject to future evaluation, this focused approach is intended to help compensate for the current gap in QIS training data, while anticipating that the system will become even more effective as more data becomes available. The figures provided utilize screen captures from Microsoft Katas \cite{mykhailova2020quantum}—a useful reference point for this work—merely to demonstrate these overarching challenges, not as a critique of the tool itself.

\subsection{Quantum Katas}

\subsubsection{Orthogonality: Lack of Contextual Awareness}

Quantum Katas \cite{mykhailova2020quantum}, although helpful, is plagued by an "orthogonality" problem where the LLM Copilot (as of March of 2025) operates independently of the learning context. This can be seen when the Copilot delivers generic, rather than lesson-specific, responses.

\begin{figure}[htbp]
    \centering
    \includegraphics[width=0.6\linewidth]{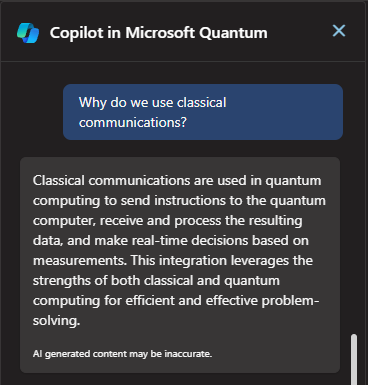}
    \caption{Lack of Context: Copilot gives a generic answer about classical communication, not specific to the quantum teleportation lesson.}
    \label{fig:katas_classical_communication}
\end{figure}

\begin{figure}[htbp]
    \centering
   \includegraphics[width=0.6\linewidth]{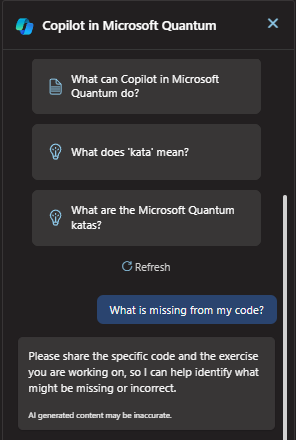}
    \caption{Missing Context: Copilot asks for code it should already have access to, highlighting system orthogonality.}
    \label{fig:katas_code_error}
\end{figure}

Figure \ref{fig:katas_classical_communication} illustrates a generic response to "Why do we use classical communication?". The Copilot does not connect the explanation to the current module on quantum teleportation. The response, while generally correct, is not specific to the student's current learning requirements. This issue is exacerbated when students request help with errors. The Copilot (specifically in the online Quantum Katas platform referenced), unaware of the student's code and assignment, simply requests the code, despite its ready availability (see Figure \ref{fig:katas_code_error}), a limitation not present in the VS Code integration.

This absence of contextualization and integration, where the various components (lesson display, code editor, AI assistant) appear to operate without shared awareness, is a fundamental characteristic of orthogonality in this setting. It impedes the Copilot's ability to provide effective, contextual support, as activity within one component fails to inform the others.


\subsubsection{Static Learning Paths: Limited Flexibility}

\begin{figure}[htbp]
    \centering
    \includegraphics[width=0.6\linewidth]{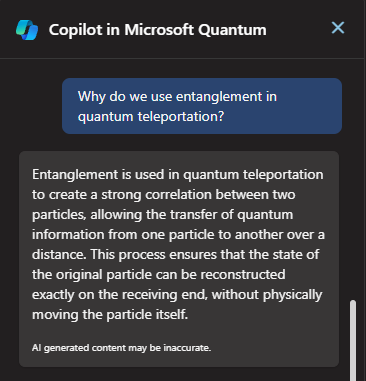}
    \caption{Static Learning Path: Copilot provides a generic explanation, not a personalized response to the student's confusion.}
    \label{fig:katas_entanglement_confusion}
\end{figure}

Quantum Katas offers a set educational route \cite{mykhailova2020quantum} --although extremely valuable compared to traditional alternatives (e.g., a book with few examples). Students move in sequence, with little options for individualized support or an alternative approach to solving the same problem(s). There is no easy way to get extra support on specific sub-skills. When confused ("Why do we use entanglement in quantum teleportation?"), the Copilot provides an accurate but generic explanation. Figure \ref{fig:katas_entanglement_confusion} shows the response without any references to the particular problem (e.g., the StandardTeleport operation within Kata 15, Teleportation \cite{mykhailova2020quantum}).

The student has few choices: {\it reread, struggle, refer to the generic Copilot, or expose the solution. There's no mechanism for targeted remediation or a sub-lesson on entanglement}. This rigidity limits the system's adaptability to various learning styles.

\subsubsection{LLM Dependence: The "Over-Reliance" Problem}

LLMs, though convenient, may become a replacement for critical thinking ("over-reliance"), inhibiting learning \cite{zhaieffects}. Students can over-depend on the LLM for answers, hence weakening their motivation to learn the content deeply. In Quantum Katas, the Copilot can sometimes provide near-complete solutions \cite{mykhailova2020quantum}.

\begin{figure}[htbp]
    \centering
    \includegraphics[width=0.6\linewidth]{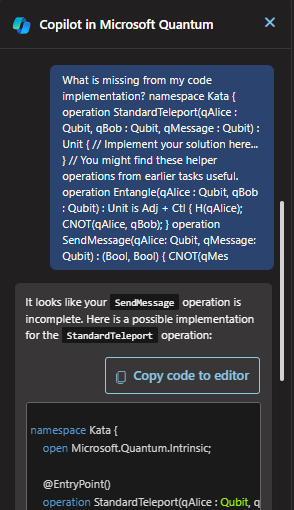}
    \caption{LLM Dependence: Copilot provides a near-complete solution, potentially hindering deep learning.}
    \label{fig:katas_full_solution}
\end{figure}

Figure \ref{fig:katas_full_solution} illustrates the Copilot producing a substantial part of the StandardTeleport solution after being given incomplete code.
This is more than hints, giving enough information to complete the task without full understanding. This fosters dependence \cite{zhaieffects}, preventing the development of authentic quantum insight and possibly leading to detrimental long-term learning outcomes \cite{bastani2024generative}.

\subsection{Addressing Shortcomings}

Our platform seeks to minimize these limitations, aiming to provide a more effective and customized learning experience, subject to further evaluation. An extensive evaluation conducted across various quantum computing topics is underway, with preliminary results anticipated for our oral presentation.

\subsubsection{Non-Orthogonality: Contextual Awareness and Integration}

We address orthogonality in our design by introducing a knowledge graph intended to tightly integrate all system components. The LLM Teaching Agent, informed by the graph, is designed to maintain a continuous understanding of the student’s state (current lesson, past interactions, code, video-lecture history). The agent always has access to a complete and accurate lesson plan.

\begin{figure}[htbp]
    \centering
   \includegraphics[width=0.69\linewidth]{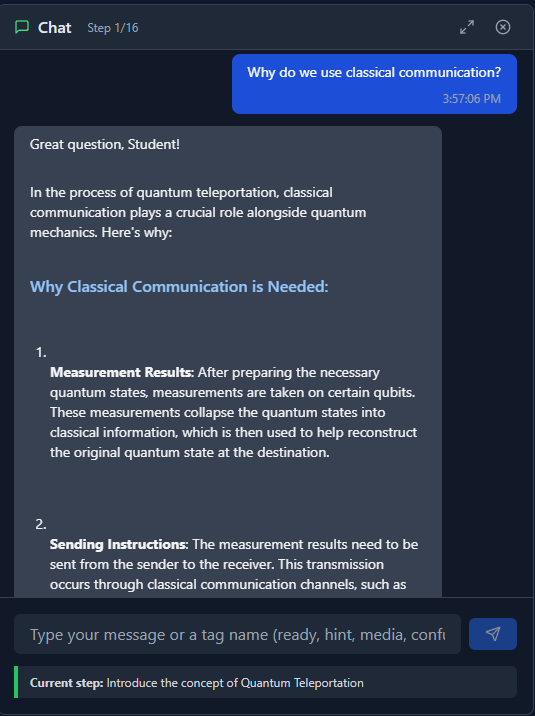}
    \caption{Contextual Awareness: The agent has access to a detailed and accurate lesson plan.}
    \label{fig:our_ortho_plan}
\end{figure}

When a student asks a question or encounters an error, the Teaching Agent is intended to tap into this rich context to offer relevant, specific support. The responses are intended to be custom-tailored to the \textbf{specific} issue, based on the student’s code, instructional objectives, and related background. The knowledge graph design ensures all parts of the system can be interconnected.

\subsubsection{Dynamic Learning Paths: Flexibility and Personalization}
\begin{figure}[htbp]
    \centering
   \includegraphics[width=0.69\linewidth]{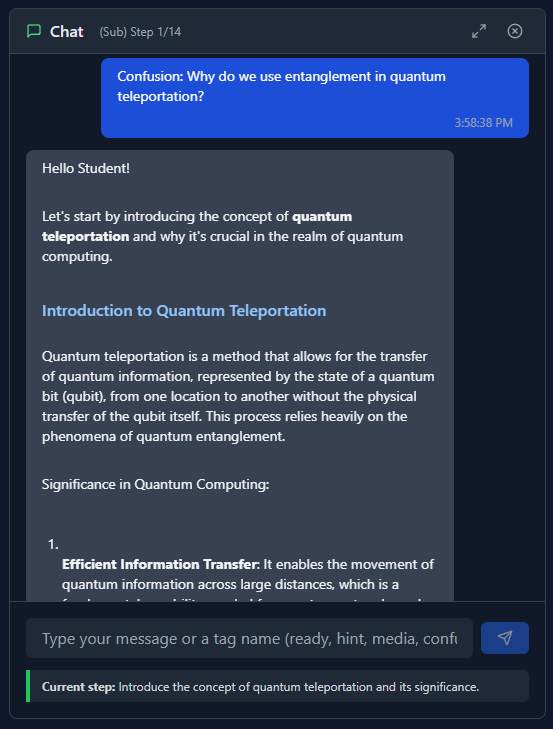}
    \caption{Dynamic Learning: "Confusion" tag triggers a personalized sub-lesson, adapting to the student's needs.}
    \label{fig:our_system_confusion}
\end{figure}

We aim to move beyond static routes by incorporating mechanisms for dynamic adaptation (see Figure \ref{fig:our_system_confusion}). We introduce the ”Confusion” tag, which is pivotal to this flexibility. Using this tag initiates the Lesson Planning Agent to create a personalized ”sublesson”.

This sub-lesson might review fundamentals, offer alternative explanations, or provide further practice. After completing the sub-lesson and demonstrating comprehension, the student returns to the main lesson. This dynamic adaptation is intended to accommodate individual learning styles and paces.

\subsubsection{Mitigating LLM Dependence: Guided Learning and Controlled Assistance}

Our system is designed to alleviate the ”over-reliance” issue \cite{zhaieffects} by carefully controlling the level of assistance provided by the LLM. The Teaching Agent is explicitly instructed \textbf{never} to provide complete solutions. Instead, it offers hints, explanations, and directions \cite{li2024tutorly}, aiming to encourage students to actively engage with the content and enabling the growth of independent problem-solving skills.

\begin{figure}[htbp]
    \centering
   \includegraphics[width=0.45\linewidth]{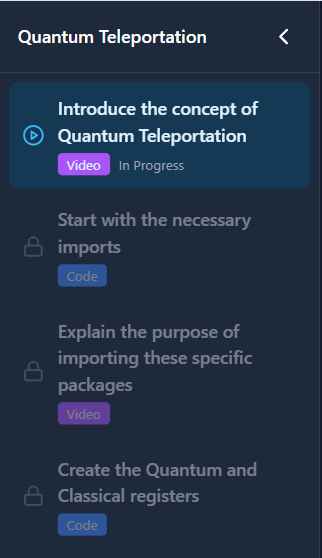}
    \caption{Guided Learning: Step-by-step lesson plans promote incremental knowledge building, reducing LLM dependence.}
    \label{fig:our_lesson_plan}
\end{figure}

The Lesson Plan Agent produces sequenced, structured lesson plans (see Figure \ref{fig:our_lesson_plan}), which are intended to enable gradual knowledge building \cite{hu2024teaching}. Instructors develop such plans, aligning with our flipped-classroom approach \cite{giannakos2014reviewing}, so that students have the necessary background for deep in-class discussion.

We anticipate this organized approach, together with focused assistance, will support fuller understanding while avoiding overreliance on the LLM. By aiming to address orthogonality, provide individualized learning paths, and limit overreliance on the LLM, our approach seeks to maintain a productive learning environment, though its effectiveness requires empirical validation \cite{chrisochoides2024developing}. Additionally, our approach encourages discussion of issues (in the class) that teachers could not predict ahead of time (as would be possible under a fixed learning path where questions can be anticipated), accommodating the diverse backgrounds of students in quantum computing courses ~\cite{wilcox_24, PhysRevPhysEducRes.18.010150}.

\section{Data-Driven LLM-Powered Approach}

The initial design was centered on integrating three basic modules \cite{chrisochoides2024developing}:

\begin{itemize}
    \item \textbf{Video Player (VP)}: Used for showing video lectures and instructional tutorials.
    \item \textbf{Code Editor (IDE)}: Utilized for writing and executing quantum code.
    \item \textbf{Chat Interface (CI)}: For student-system interaction.
\end{itemize}

\begin{figure}[htbp]
    \centering
    \includegraphics[width=0.8\linewidth]{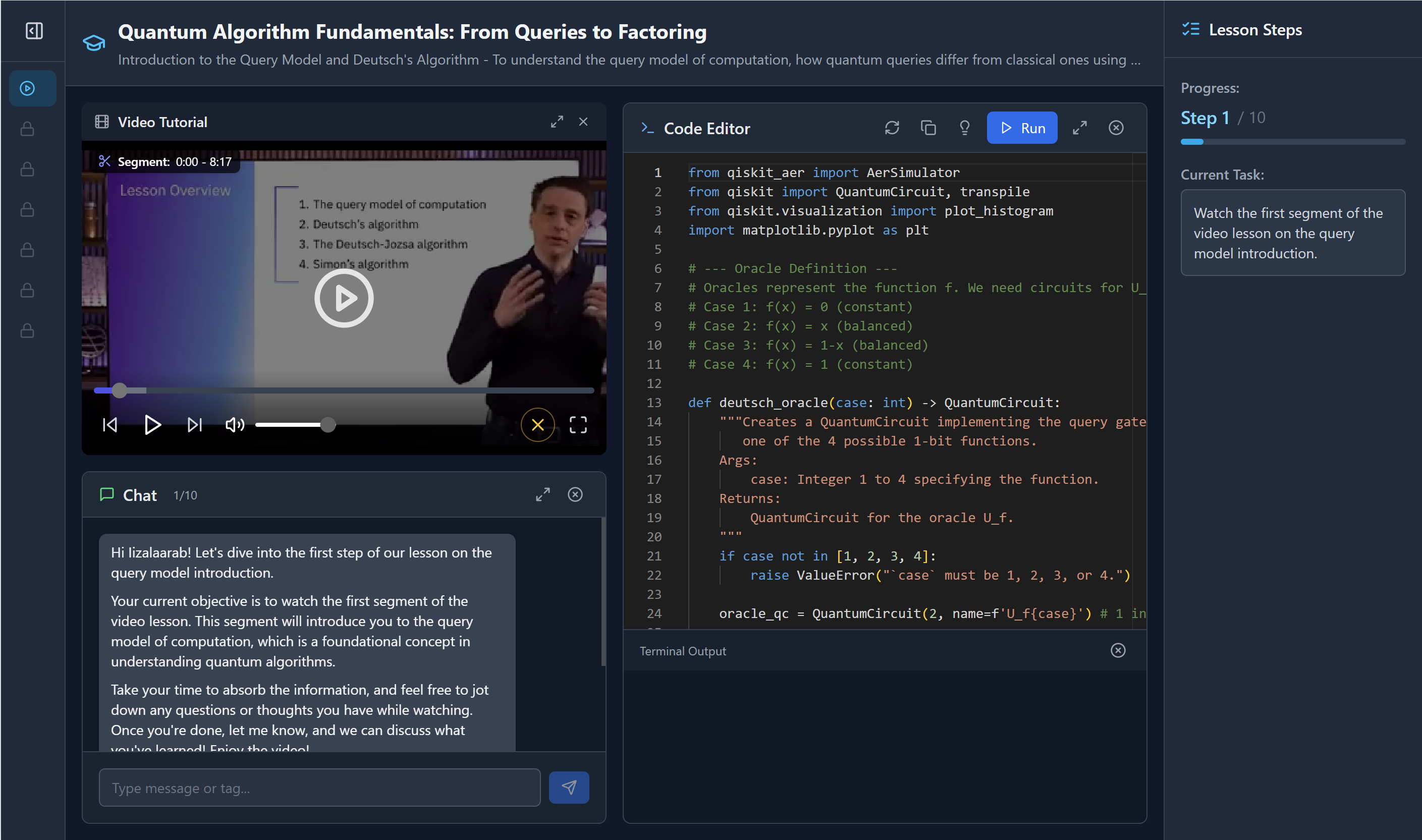}
    \caption{ITAS interface during Step 1/10 of the "Quantum Algorithm Fundamentals" lesson\cite{quantumQueryAlgorithms2023}, showing integrated video, code editor, and chat components.}
    \label{fig:initial_design}
\end{figure}

Our goal is to create a unified learning environment that supports effortless switching among various learning resources. Importantly, we aim to move away from traditional linear learning paths with bounded adaptability (see Section III.A).
We built a unique Large Language Model (LLM) augmented with custom logic to accomplish this goal and aim to provide support that goes beyond accuracy alone, potentially offering better adaptability and understanding. Specifically, the "Augmented LLM" is architected to:
\begin{itemize}
    \item \underline{Develop Lessons}: Create thoughtful and systematic instructional lessons based on educator input (e.g., "Teach quantum teleportation using this video and code example").
    \item \underline{Facilitate Learning}: Apply the established instructional lessons, providing explanations, assistance, and assessments to the student.
    \item \underline{Regulating Progress}: Determine the proper point at which the student is ready to move on to the next stage in the lesson.
    \item \underline{Call Tools}: Interact with various parts of the system, such as the code editor and video player, by using predefined tools (e.g., "seek to timestamp 2:30 in the video," "highlight line 5 in the code editor").
\end{itemize}

\begin{figure}[htbp]
    \centering
    \includegraphics[width=0.89\linewidth]{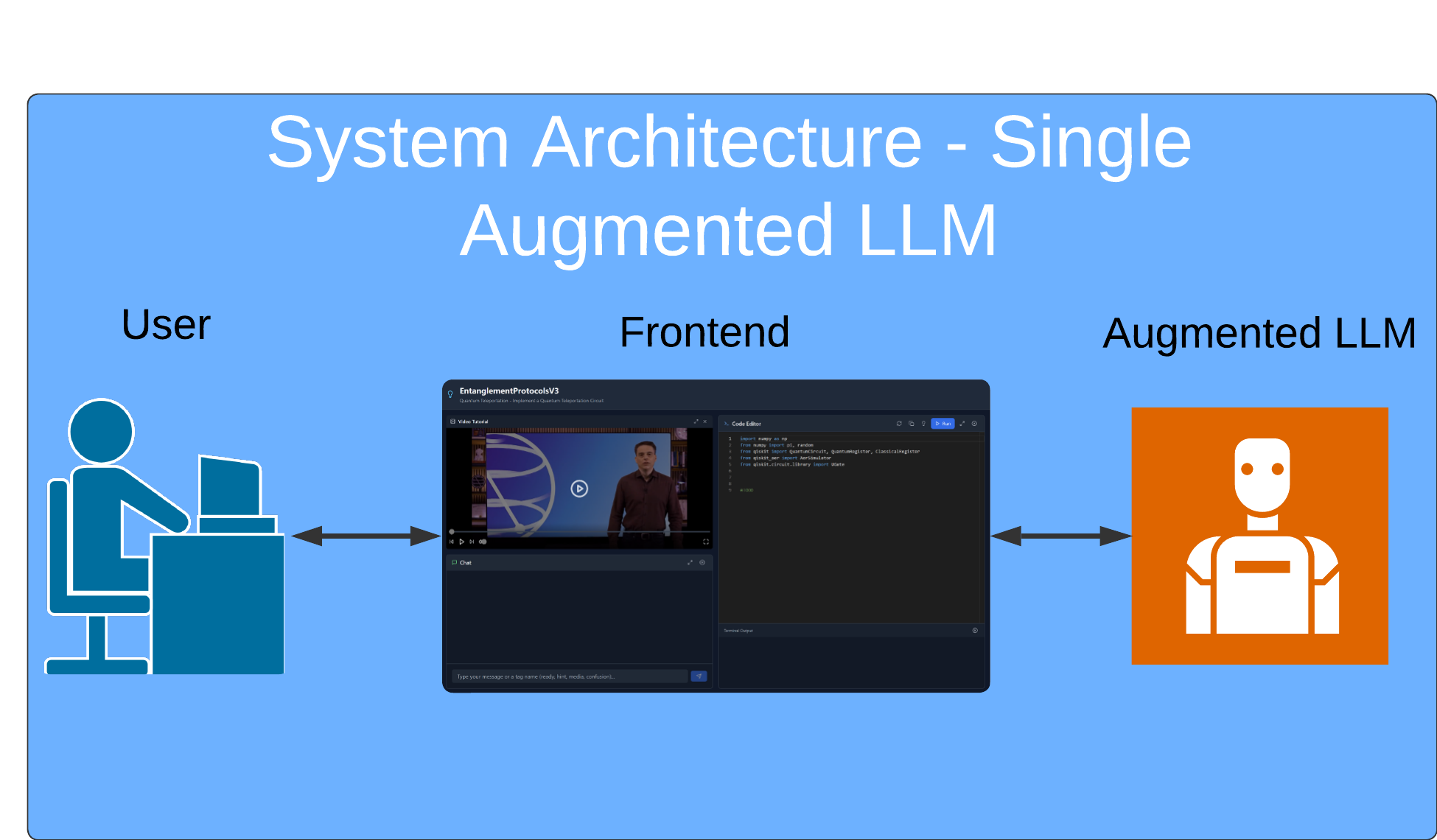}
    \caption{Single Augmented LLM with Planning, Teaching, Pacing, and Tool-Calling Capabilities}
    \label{fig:first_iteration}
\end{figure}

This initial single-agent approach aimed to simplify and ensure the consistent and tight integration of VP, IDE, and CI modules (or system modules). The LLM has access to all three modules and is responsible for overseeing all aspects of the instructional process.

\begin{figure}[htbp]
    \centering
    \includegraphics[width=0.69\linewidth]{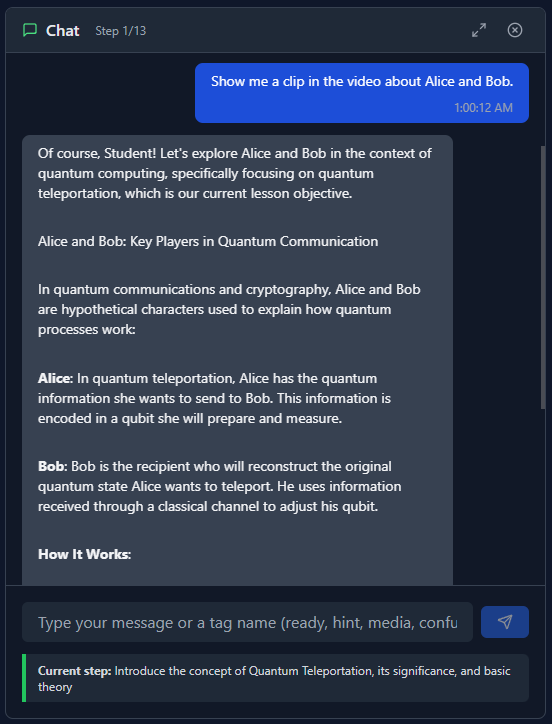}
    \caption{Example of LLM Hallucination: Incorrect Tool Usage. Despite the user requesting a video clip, the system provides a textual explanation instead of using the media player.}
    \label{fig:hallucination_example}
\end{figure}

While promising, the single Augmented LLM quickly encountered significant challenges:

\begin{itemize}
    \item \underline{Hallucinations}: The language model had a tendency to hallucinate, generating incorrect tool calls (e.g., calling unnecessary tools or not calling required ones), providing inaccurate explanations, and deviating from the lesson plan, as shown in Figure [\ref{fig:hallucination_example}].
    \item \underline{Inadequate Pacing Choices}: The large language model had trouble assessing student understanding accurately and often moved to the next lesson step too early or too late, as presented in Figure [\ref{fig:hallucination_example}].
    \item \underline{Overload}: The combination of planning, teaching, pacing, and tool-calling responsibilities overloaded the LLM, contributing to the hallucination problem, see Figure [\ref{fig:hallucination_example}].
\end{itemize}

These issues significantly degraded the learning experience and highlighted the limitations of a single, overloaded LLM.

\paragraph*{\bf Separation of Concerns - Two-Agent Architecture}

To alleviate the problems of hallucination and information overload, we applied the principle of "separation of concerns" \cite{lecun2022path}, dividing the tasks between two specialized LLM agents \cite{wang2025llm, chu2025llm}:

\begin{itemize}
    \item \underline{Lesson Planning Agent}: Responsible for generating and revising lesson plans \cite{hu2024teaching}. It considers instructor input, including the subject of the lesson, learning objectives, and available resources, to produce a coherent, ordered lesson plan. It is also triggered by a student's indication of confusion, creating "sub-lessons" to address specific knowledge gaps.
    \item \underline{Teaching Agent}: It is responsible for carrying out the lesson plan formulated by the Lesson Planning Agent. It provides explanations, recommendations, and evaluations \cite{li2024tutorly}, and interacts with the student through the chat interface.Crucially, in this version, the Teaching Agent had the responsibility of moving to the next stage in the lesson plan when it deemed necessary.
\end{itemize}

\begin{figure}[htbp]
    \centering
    \includegraphics[width=0.89\linewidth]{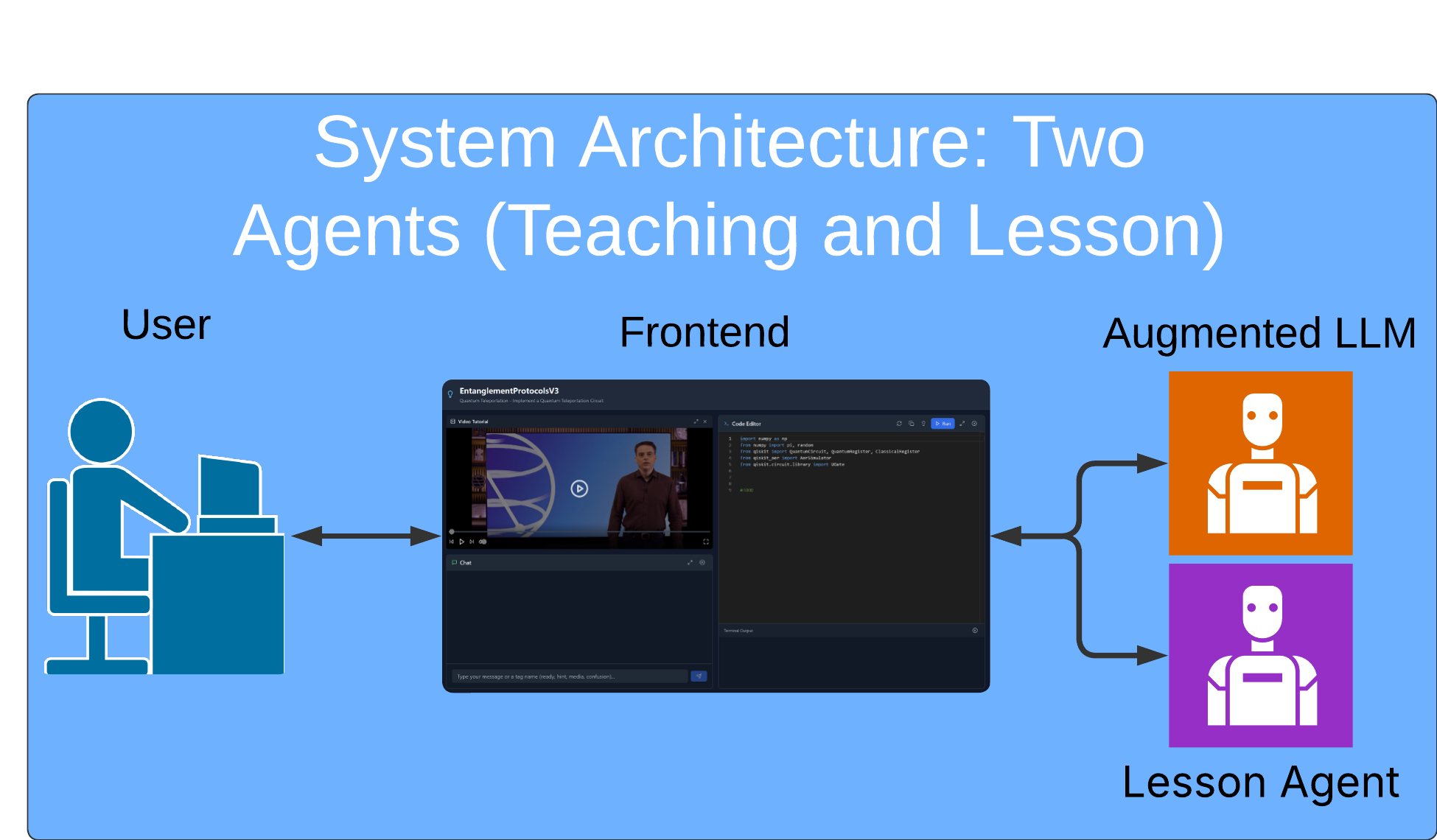}
    \caption{Two-Agent Architecture with Separate Lesson Planning and Teaching Agents.}
    \label{fig:second_iteration}
\end{figure}

This two-agent architecture significantly reduced the burden on each LLM, improving performance and reducing hallucination. Despite this, a new challenge was posed: {\it Communication and State Management}.
Implementation of this dual-agent system significantly eased the load on each of these LLMs, resulting in improved performance with a decrease in occurrences of hallucination. However, demarcating roles between a Lesson Planning Agent and a Teaching Agent immediately highlights a critical requirement: an efficient communication channel with a shared knowledge state \cite{chu2025llm}. To say it in simpler words, without some means for these agents to exchange information and access a common understanding of the learning environment, the system is rendered inoperable. This concept is a natural extension of the 'separation of concerns' principle; just as separate components require well-defined interfaces, individual agents require some shared knowledge base.

\paragraph*{\bf The Knowledge Graph}

To address the requirement of enabling communication and coordinated activity, we introduced a Knowledge Graph \cite{qu2024survey, ain2024learner}. The Knowledge Graph is designed to serve not merely as a memory store, but as the critical foundation upon which the Lesson Planning Agent and Teaching Agent can potentially collaborate effectively \cite{jaldi2025education}, pending further system integration and testing. Without it, they simply could not carry out their respective tasks, rendering the entire architecture non-functional. More specifically, the Knowledge Graph acts as a central, persistent, and structured repository that stores:

\begin{itemize}
    \item \underline{Users}: Students and instructors.
    \item \underline{Learning Resources}: Videos, code, lesson plans.
    \item \underline{Interactions}: Student activities (e.g., viewing a video segment, submitting code, asking a question), system responses (e.g., providing a hint, displaying an explanation), and agent activities (e.g., generating a lesson plan, modifying a step).
    \item \underline{Student State}: A continuously updated representation of the student's knowledge and interaction history, including viewed resources, code submissions, and chat history.
\end{itemize}

\begin{figure}[htbp]
    \centering
    \includegraphics[width=0.79\linewidth]{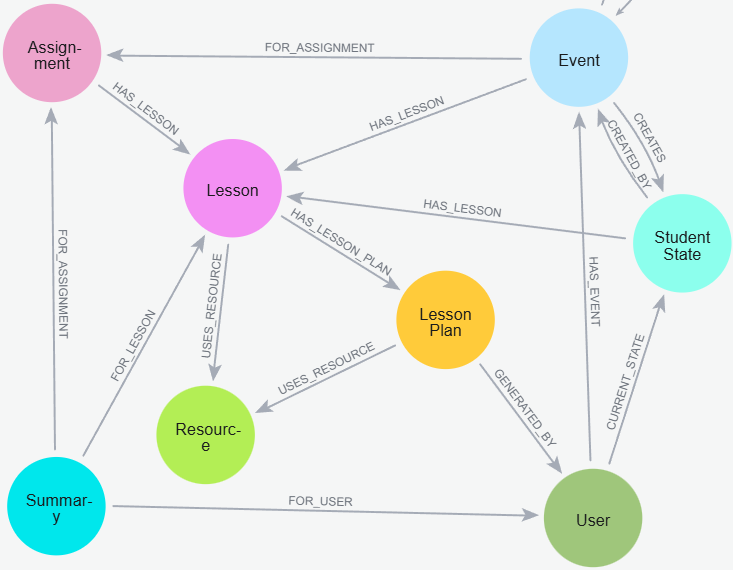} 
    \caption{Partial View of the Graph Data Model. Different colors represent different node types. Arrows indicate relationships and their directions.}
    \label{fig:graph_data_model}
\end{figure}

The knowledge graph is designed to facilitate:

\begin{itemize}
    \item \textbf{Context-Aware Tutoring}: The Teaching Agent can access the student's past interactions and ongoing status to provide personalized feedback and suggestions \cite{li2024tutorly}.
    \item \textbf{Agent Communication}: The Lesson Planning Agent can update the lesson plan in the graph, and the Teaching Agent can retrieve the new plan \cite{wang2025llm}.
    \item \textbf{Insights Derived from Data}: The combined interaction data included in the graph provide a basis for future learning path analysis and enhancement \cite{giannakos2014collecting}.
\end{itemize}

\begin{figure}[htbp]
    \centering
    \includegraphics[width=0.69\linewidth]{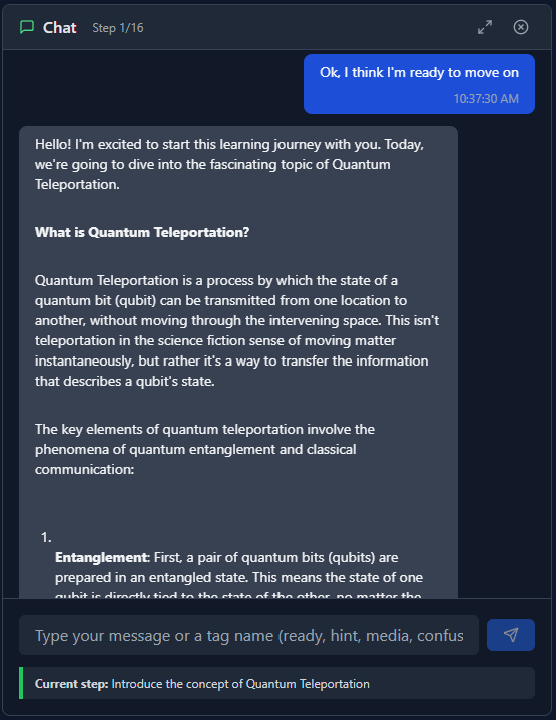}
    \caption{Example of Teaching Agent Hallucination: Failure to Recognize User Intent. The system fails to advance to the next step despite the student explicitly expressing their intent.}
    \label{fig:teaching_agent_hallucination}
\end{figure}

The KG provided shared state and communication, enabling the agents to work together effectively. Nevertheless, even with this essential component present, and despite the two-agent system, the Teaching Agent continued to manifest hallucinations, specifically when attempting to determine the appropriate timing for advancing to the next step of a lesson. The agent, which still had the responsibility of determining student understanding, sometimes made incorrect pacing choices, e.g., failing to move on to the next step even when the student explicitly indicated readiness. One example of this deficiency is shown in Figure \ref{fig:teaching_agent_hallucination}. In this case, the student has explicitly stated, 'Ok, I think I'm ready to move on,' yet the Teaching Agent is still in the current step, continuing to introduce the concept of quantum teleportation instead of proceeding as requested. This highlighted a fundamental weakness of LLMs: their inability to infer user intent from natural language input reliably, even with seemingly obvious requests (Figure \ref{fig:teaching_agent_hallucination}).

\paragraph*{\bf  User-Driven Intent - The Tag System}

To address the matter of hallucinations and increase user agency, we introduced a \textbf{Tag System}. Instead of relying on the Teaching Agent to infer student readiness, we empowered students to explicitly state their intentions through designated tags.

\begin{figure}[htbp]
    \centering
     \includegraphics[width=0.99\linewidth]{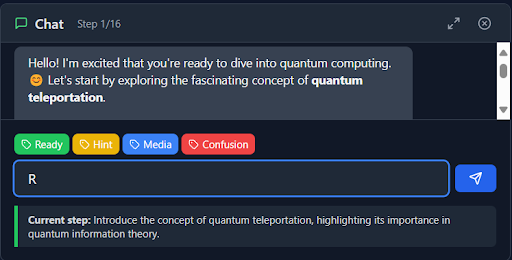}
     \caption{Tag Selection Interface.}
    \label{fig:chat_tags}
\end{figure}

Key tags include:

\begin{itemize}
    \item {\it Ready}: Signals that the student is prepared to proceed to the next step of the lesson.
    \item {\it Hint}: Requests a hint for the current programming task.
    \item {\it Media}: Requests a video content search.
    \item {\it Confusion}: Expresses confusion and triggers the Lesson Planning Agent to generate a sub-lesson.
\end{itemize}

The tag system is intended to achieve several crucial goals:

\begin{itemize}
    \item \textbf{Aims to Reduce Hallucination}: By removing the need for the Teaching Agent to infer student intention, we significantly reduce the likelihood of hallucination.
    \item \textbf{Enhance User Control}: Students have direct control over the speed of the instructional session and can ask for particular kinds of help.
    \item \textbf{Provides Adaptability}: The tagging model is made extensible to allow the addition of new tags to cater to additional student needs and increase system functionality.
    \item \textbf{Offers Structure Input}: The use of tags offers structured input to the LLMs, which is far simpler for them to process than unstructured natural language.
\end{itemize}

The implementation of the tag system is a significant step ahead in the development of our system, towards a more solid, dependable, and learner-oriented learning experience.

\vspace{0.5cm}
\paragraph*{\bf Putting It All Together}

The current system architecture combines the following:

\begin{itemize}
    \item \textbf{Two Specialized LLM Agents}: Lesson Planning Agent and Teaching Agent, each with clearly defined responsibilities \cite{wang2025llm}.
    \item \textbf{Knowledge Graph}: A central, persistent memory that stores all system data and enables communication and state management \cite{skjaeveland2024ecosystem, qu2024survey}.
    \item \textbf{Tag System}: An interface framework for enabling students to articulate their intentions and manage the learning process.
    \item \textbf{Frontend Components}: Video player, code editor, chat interface, and lesson presentation.
\end{itemize}

This system is designed to directly address the key shortcomings (subject to extensive future evaluation) identified in existing quantum education platforms: orthogonality, reliance on static learning flows, over-reliance on LLMs, and the ubiquitous problem of LLM hallucination. By incorporating a two-agent system, a central Knowledge Graph, and a user-driven tag system, the resultant platform provides a much more personalized, adaptive, and robust learning experience for quantum computing education.

\section{Preliminary Results}

To collect initial data and assess the central functions of our ITAS, we carried out initial tests that mimicked a student accessing a particular lesson module: \textbf{\"Quantum Algorithm Fundamentals: From Queries to Factoring\"}\cite{quantumQueryAlgorithms2023}. This lesson introduces the Query Model of computation and Deutsch's Algorithm, and explains how quantum queries differ from classical ones through unitary gates. It then guides students through implementing their first quantum algorithm (Deutsch's) and a quantum advantage. The content and structure come from Lesson 5 ("Quantum Query Algorithms") of John Watrous' "Fundamentals of Quantum Algorithms" Qiskit textbook material \cite{quantumQueryAlgorithms2023}, translated to our interactive platform. Figure~\ref{fig:initial_design} presents the ITAS interface in Step 1 of this learning module, with the video tutorial, code editor, and chat interface.


\subsection{Data Collection and Context Awareness}
During these simulated sessions, we captured a broad set of interaction events from the system. As explained in Table~\ref{tab:lesson_node_counts_compact}, these events originate from all system components: the video player (e.g., play/pause, seeks, heartbeats related to video engagement), the code editor (e.g., successful code submissions), and the chat interface (e.g., user messages, agent actions, tag usage, tool calls).

\begin{table}[htbp]
    \centering
    \caption{Node and event counts for a single simulated lesson run.}
    \label{tab:lesson_node_counts_compact}
    \small 
    \setlength{\tabcolsep}{4pt} 
    \renewcommand{\arraystretch}{0.95} 
    \begin{tabular}{@{}lr@{}}
        \toprule
        \textbf{Category / Type}                 & \textbf{Count} \\
        \midrule
        \multicolumn{2}{l}{\textbf{Core Entities}} \\
        User / AI Assistant         & 1 / 1 \\ 
        Assignment / Lesson         & 1 / 1 \\ 
        Lesson Steps              & 15    \\ 
        Summary                   & 1     \\
        \midrule
        \multicolumn{2}{l}{\textbf{Events}} \\
        Video Heartbeat           & 256   \\
        Chat Assistant            & 29    \\
        Video Seeked                & 15    \\
        Step Completed             & 15    \\ 
        Chat User                 & 15    \\
        Video Play / Pause          & 7 / 7 \\ 
        Segment Req / End         & 3 / 3 \\ 
        Video Vol Change          & 3     \\ 
        Segment Reset               & 1     \\
        Lesson Start / End        & 1 / 1 \\ 
        Code Success                & 1     \\ 
        Chat Tool Call            & 1     \\ 
        Metadata Load             & 1     \\ 
        \midrule
        \multicolumn{2}{l}{\textbf{Totals}} \\
        Total Core Entities       & 20    \\
        Total Events                & 359   \\
        \midrule
        \textbf{Grand Total Nodes}      & \textbf{379} \\
        \bottomrule
    \end{tabular}
    \renewcommand{\arraystretch}{1.0} 
\end{table}

All of this interaction data, combined with the lesson structure (represented as nodes and edges) and the student's progress state, is housed within the central \textbf{Knowledge Graph}. This unified data representation makes contextual information readily accessible to both the \textbf{Teaching Agent} and the \textbf{Lesson Planning Agent}. This directly confronts the issue of \textbf{"orthogonality"} present in systems where individual components operate (see Section III.A), as agents can draw on past information (video watched, code tried, past questions) from all sources to provide individualized assistance.

The data gathered will be mined in later analyses to extract insights into student learning behaviors, determine effective pedagogical approaches, and establish best teaching practices, thus improving content quality and teaching efficacy across quantum computing subjects.

\subsection{Dynamic Lesson Adaptation}
One of the important aspects illustrated in this session is the system's capability of \textbf{dynamic adjustment in real-time} within this context. In the lesson, the student activated the \textbf{"Confusion"} tag, which explicitly indicated the requirement for clarification (similar to the conceptual example in Section III.B).

\begin{figure}[htbp]
    \centering
    \includegraphics[width=0.95\linewidth]{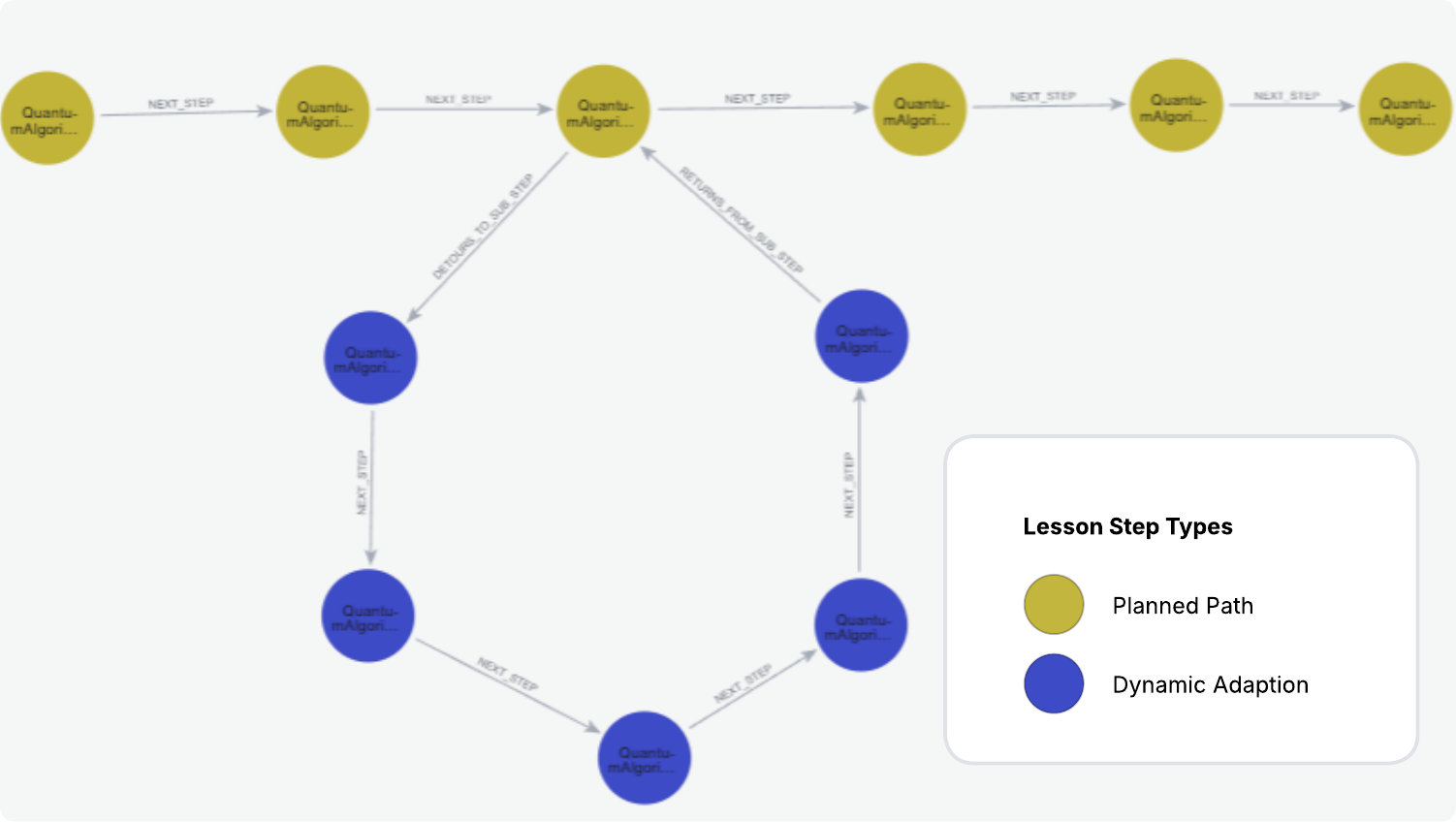}
    \caption{Lesson plan graph structure after the "Confusion" tag triggered the insertion of a sub-lesson (represented by the hexagonal loop/diversion), demonstrating dynamic adaptation based on student feedback.}
    \label{fig:lesson_plan_after_confusion}
\end{figure}

This explicit request invoked the \textbf{Lesson Planning Agent}. The preliminary lesson plan followed a relatively linear sequence of steps. However, upon sensing the "Confusion" tag, the agent dynamically revised the current lesson plan by inserting a targeted \textbf{"sub-lesson"} to clarify the specific source of confusion. Figure~\ref{fig:lesson_plan_after_confusion} illustrates the conceptual structure of the lesson plan \textbf{after} this dynamic revision, showing a detour (the cycle involving "RETURNS\_FROM\_SUB\_STEP" edges) from the main path to cover the prerequisite material or offer clarification before returning the student to the main flow.

This contrasts sharply with the static learning paths common in many platforms. Our system demonstrates the potential to enable the learning experience to evolve dynamically in response to explicit student feedback, supported by the \textbf{tag system} and the generative potential of the \textbf{Lesson Planning Agent} operating on the \textbf{Knowledge Graph representation} of the lesson.

\begin{figure}[htbp]
    \centering
    \includegraphics[width=0.6\linewidth]{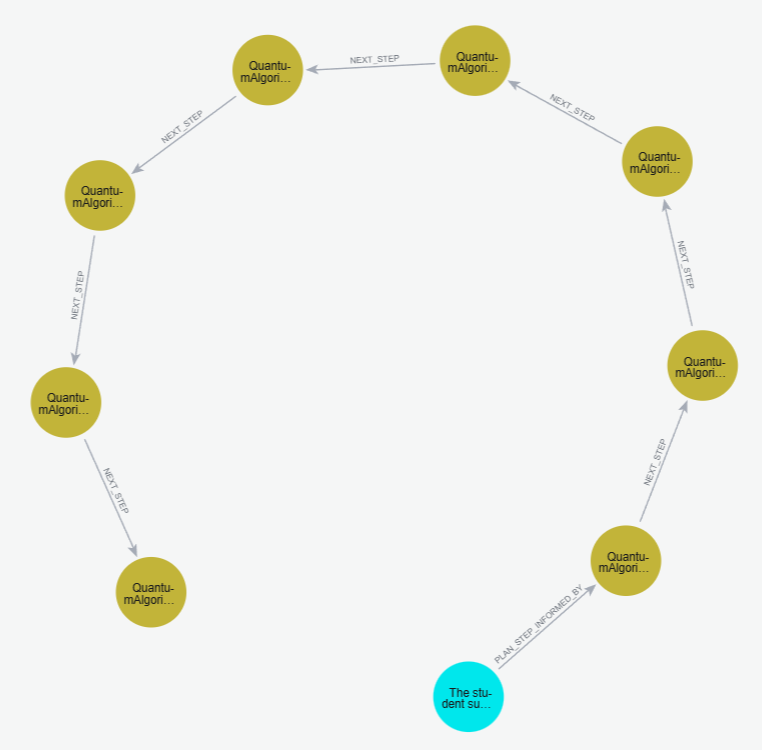}
    \caption{Knowledge graph excerpt illustrating how a summary node (bottom, cyan), potentially derived from a \textbf{different student's} previous interaction state, informs the planning of a subsequent lesson segment (relationship labeled "PLAN\_STEP\_INFORMED\_BY").}
    \label{fig:summary_informed_planning}
\end{figure}

Furthermore, the system design incorporates mechanisms intended for \textbf{continuity}, as illustrated by the summary node concept. Following the conclusion of a lesson or major portion, a summary node can be generated in the Knowledge Graph. At the start of a \textbf{new} lesson plan, the Lesson Planning Agent can use such summaries as input context. Figure~\ref{fig:summary_informed_planning} illustrates how a node representing a previous learning session(bottom, cyan) assists in the generation of a subsequent learning plan. This design aims to ensure that subsequent interactions can benefit from previous experience, preserving context over time.

Later versions will attempt to automate these real-time adaptations by inferring student status, confusion, and intent directly from interaction patterns instead of explicitly selecting tags.

\subsection{Video Interaction Analysis and Data Mining Potential} 
The granularity of the collected video interaction data (e.g., 256 \textbf{'Video Heartbeat'} events depicted in Table~\ref{tab:lesson_node_counts_compact}, in addition to seeks, plays, and pauses) stored within the Knowledge Graph presents opportunities for data mining and the acquisition of insights into patterns of student engagement \cite{giannakos2014collecting, giannakos2015making}. Rather than merely determining whether a video was watched, it is possible to investigate \textbf{how} it was watched.

\begin{figure}[htbp]
    \centering
    \includegraphics[width=0.75\linewidth]{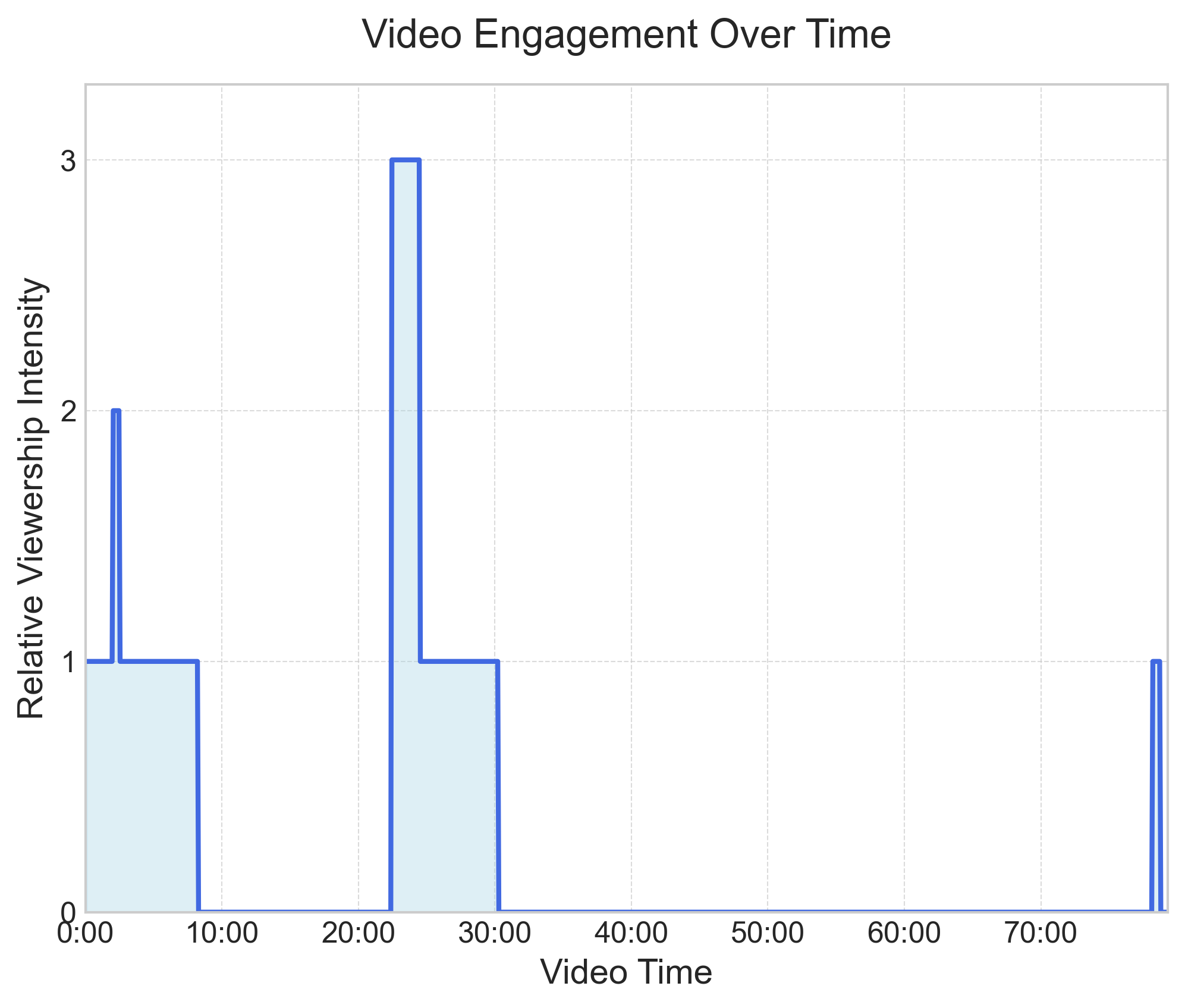}
    \caption{Simulated video engagement intensity over time, derived from video interaction events. Peaks suggest re-watching or high interest, while gaps indicate skipped sections.}
    \label{fig:video_engagement_graph}
\end{figure}

Figure~\ref{fig:video_engagement_graph} provides an example visualization derived from such data, plotting relative viewership intensity over the video's duration. This type of analysis can reveal:
\begin{itemize}
    \item \textbf{Peaks in Engagement:} Areas that were visited numerous times (where intensity \textgreater 1) may be associated with challenging concepts needing more investigation or subjects of great student interest. The high peaks observed at the 1-minute and 23-minute marks indicate possible areas of re-viewing or concentrated attention.
    \item \textbf{Plateaus:} Periods of steady viewing (intensity = 1) indicate continuous engagement.
    \item \textbf{Gaps or Low Intensity:} Periods with no recorded heartbeats or activity might indicate skipped sections or student disengagement. The long gap between roughly 30 minutes and 75 minutes clearly shows a lack of viewership in that segment during this simulated run.
\end{itemize}

Analyzing patterns, especially when aggregated across groups of students, can provide instructors with valuable feedback about which parts of a video lecture are effective or challenging. It also lays the groundwork for \textbf{learnersourcing} techniques \cite{kim2015learnersourcing}, in which collective interaction data could be used to automatically identify important segments, suggest navigation points, or even trigger proactive interventions by the Teaching Agent if patterns indicative of confusion are detected across many users. While this preliminary run only shows data from a limited number of simulations, it demonstrates the system's capacity to capture the necessary data for such future analyses.

These initial suggest the potential of our approach in: (1) gathering rich, multi-modal interaction data within a knowledge graph, exhibiting \textbf{non-orthogonality}, (2) using this graph, along with specialized agents and a user-friendly tagging feature, to facilitate \textbf{context-aware}, \textbf{dynamically adaptive} learning experiences (as demonstrated in simulation), and (3) collecting fine-grained data potentially suitable for an in-depth examination of student activity, especially with video content, showcasing the potential for \textbf{data-driven insights}. Further validation with real users is needed to confirm these capabilities.

\section{Conclusion and Future Work}


This paper presented a novel intelligent tutoring system for quantum computing education, highlighting its innovative design process. Our system is designed to address significant shortcomings of current platforms, such as orthogonality, strict learning pathways, and dependence on LLMs, by incorporating a two-agent architecture, a knowledge graph for control and state sharing, and an end-user facing tag system for intent expression. The developmental process, driven by the need to overcome challenges like LLM hallucination and ensure a consistent, personalized learning experience, yielded a system intended to be robust and flexible. The definitive division of labor between a Lesson Planning Agent and a Teaching Agent, combined with the structured input provided by the tagging system, is hypothesized to significantly improve the effectiveness of LLM-powered tutoring. The knowledge graph is designed to facilitate communication and state management and may provide possibilities for future data-driven improvements and investigation of learning pathways.

Future work will involve the systematic validation of the ITAS's pedagogical efficacy, establishing best practices and benchmarks for AI tutors in quantum education, perhaps drawing on models like EAIRA \cite{cappello2025eaira}. Meanwhile, we will apply the knowledge graph for enhanced student state inference \cite{ain2024learner}, allowing for real-time learning path adjustment based on extracted data \cite{giannakos2015making}. This data-driven methodology will also inform system capability development, automation of content mining, and content improvement to continuously optimize the personal learning experience.

\section*{Acknowledgments}
This research was sponsored in part by the Richard T. Cheng Endowment. Gemini and Grammarly were used to improve readability throughout the text; the authors reviewed and take full responsibility for the final content.

\bibliographystyle{IEEEtran}
\bibliography{references}
\end{document}